\documentclass[PRB,twocolumn,nofootinbib,superscriptaddress]{revtex4-1}
\usepackage{graphicx}% Include figure files
\usepackage{dcolumn}% Align table columns on decimal point
\usepackage{bm}% bold math
\usepackage{tabularx}
\usepackage{times}
\usepackage{xcolor}
\newcommand{\crte}{CrTe$_3$}

\begin{document}

\title{Antiferromagnetism in the van der Waals layered spin-lozenge semiconductor CrTe$_3$}

\author{Michael A. McGuire}
\email{McGuireMA@ornl.gov \\  \\ Notice: This manuscript has been authored by UT-Battelle, LLC under Contract No. DE-AC05-00OR22725 with the U.S. Department of Energy. The United States Government retains and the publisher, by accepting the article for publication, acknowledges that the United States Government retains a non-exclusive, paid-up, irrevocable, world-wide license to publish or reproduce the published form of this manuscript, or allow others to do so, for United States Government purposes. The Department of Energy will provide public access to these results of federally sponsored research in accordance with the DOE Public Access Plan(http://energy.gov/downloads/doe-public-access-plan). }
\affiliation{Materials Science and Technology Division, Oak Ridge National Laboratory, Oak Ridge, Tennessee 37831 USA}
\author{V. Ovidiu Garlea}
\affiliation{Quantum Condensed Matter Division, Oak Ridge National Laboratory, Oak Ridge, Tennessee 37831 USA}
\author{Santosh KC}
\affiliation{Materials Science and Technology Division, Oak Ridge National Laboratory, Oak Ridge, Tennessee 37831 USA}
\author{Valentino R. Cooper}
\affiliation{Materials Science and Technology Division, Oak Ridge National Laboratory, Oak Ridge, Tennessee 37831 USA}
\author{Jiaqiang Yan}
\affiliation{Materials Science and Technology Division, Oak Ridge National Laboratory, Oak Ridge, Tennessee 37831 USA}
\author{Huibo Cao}
\affiliation{Quantum Condensed Matter Division, Oak Ridge National Laboratory, Oak Ridge, Tennessee 37831 USA}
\author{Brian C. Sales}
\affiliation{Materials Science and Technology Division, Oak Ridge National Laboratory, Oak Ridge, Tennessee 37831 USA}

\begin{abstract}
The crystallographic, magnetic, and transport properties of the van der Waals bonded, layered compound \crte\ have been investigated on single crystal and polycrystalline materials. The crystal structure contains layers made up of lozenge shaped Cr$_4$ tetramers. Electrical resistivity measurements show the crystals to be semiconducting, with a temperature dependence consistent with a band gap of 0.3\,eV. The magnetic susceptibility exhibits a broad maximum near 300\,K characteristic of low dimensional magnetic systems. Weak anomalies are observed in the susceptibility and heat capacity near 55\,K, and single crystal neutron diffraction reveals the onset of long range antiferromagnetic order at this temperature. Strongly dispersive spin-waves are observed in the ordered state. Significant magneto-elastic coupling is indicated by the anomalous temperature dependence of the lattice parameters and is evident in structural optimization in van der Waals density functional theory calculations for different magnetic configurations. The cleavability of the compound is apparent from its handling and is confirmed by first principles calculations, which predict a cleavage energy 0.5\,J/m$^2$, similar to graphite. Based on these results, \crte\ is identified as a promising compound for studies of low dimensional magnetism in bulk crystals as well as magnetic order in monolayer materials and van der Waals heterostructures.
\end{abstract}

\maketitle

\section{Introduction}

Electronic peculiarities and proposed applications of graphene \cite{Castro-Neto-2009, Novoselov-2012} ignited intense research into two-dimensional materials composed of layers held together by van der Waals bonding. Efforts to extend this research to other materials led, for example, to the discovery of interesting and potentially useful optoelectronic behaviors in ultra-thin transition metal dichalcogenides \cite{Splendiani-2010, Radisavljevic-2011, Xiao-2012, Xu-2014}, providing further demonstrations of new physics emerging as bulk materials are reduced to monolayers. Interest has also risen in exploring van der Waals heterostructures \cite{Geim-2013}, which requires a library of layered materials that can be mechanically exfoliated and recombined in customized stacks. The incorporation of materials with complementary functionalities, including magnetism, holds promise for the development of new physics and devices that can be controlled by external fields. While the set of two-dimensional materials beyond graphene and transition metal dichalcogenides continues to expand \cite{Lebegue-2013, Bhimanapati-2015, Ajayan-2016}, there are still relatively few studies of easily cleavable magnetic compounds and interest in such materials is growing.

Recently, several investigations of van der Waals layered ferromagnets have been reported, including CrSiTe$_3$, CrGeTe$_3$, and CrSnTe$_3$ \cite{Lebegue-2013, Li-2014, Casto-2015, Sivadas-2015, Zhuang-2015}, Fe$_3$GeTe$_2$ \cite{Deiseroth-2006, Chen-2013, May-2016}, and chromium trihalides \cite{Wang-2011, McGuire-2015, Zhang-2015, Liu-2016, Wang-2016}. Decades ago the MPS$_3$ and MPSe$_3$ families of compounds (M = divalent transition metal) were identified as van der Waals layered antiferromagnets \cite{LeFlem-1982}, and they have been revisited recently in light of their potential for studying monolayer materials \cite{Kuo-2016, Wang-2016b, Chittari-2016}. Although antiferromagnetic materials are in some ways less ``functional'' than ferromagnetic materials, interest in them is growing as ideas for their use as robust spintronic materials develop \cite{MacDonald-2011, Gomonay-2014, Jungwirth-2016}. In the present manuscript we report the results of our recent investigation of the layered antiferromagnet \crte.

\begin{figure}
\begin{center}
\includegraphics[width=3.25in]{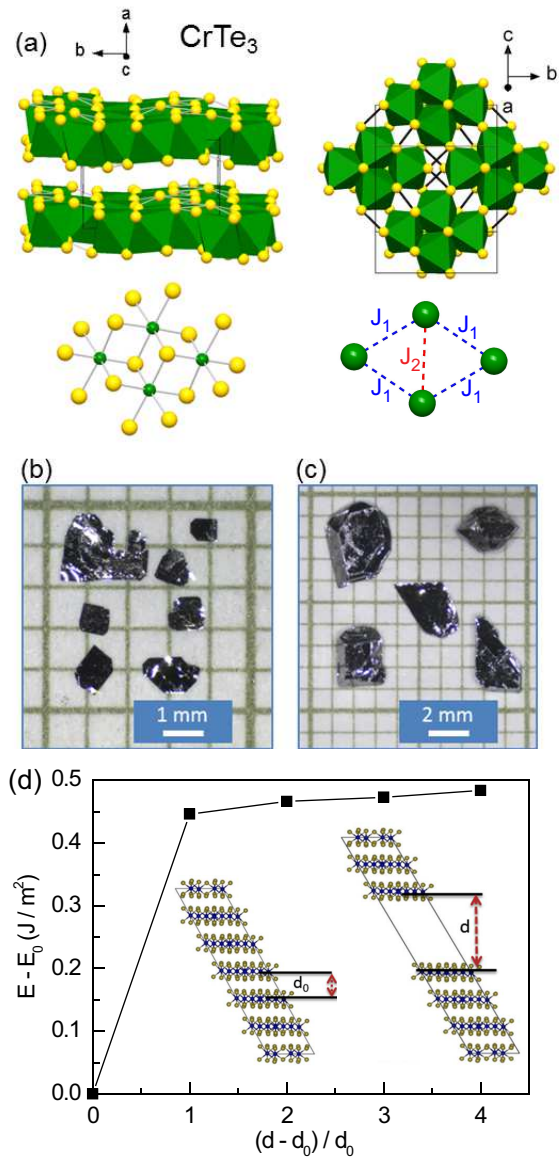}
\caption{\label{fig:crystal}
(a) The crystal structure of CrTe$_3$, showing the layers stacked along the \textit{a}-axis that are formed by corner and edge sharing Cr-centered octahedra (green) with Te atoms (yellow) at the vertices. The dark lines in the plan view of the single layer represent Te-Te bonds forming dimers and trimers. Also shown are a single Cr$_4$Te$_{16}$ unit and the lozenge shaped Cr$_4$ tetramer with magnetic exchange interactions labeled by J$_1$ and J$_2$. (b) Single crystals of CrTe$_3$ grown from an AlCl$_3$-KCl eutectic melt. (c) Single crystals of CrTe$_3$ grown from excess Te. (d) DFT-calculated energy as a function of layer separation in \crte\, indicating a cleavage energy of 0.5\,J/m$^2$.
}
\end{center}
\end{figure}

The compound \crte\ was first reported by Klepp and Ipser \cite{Klepp-1979, Klepp-1982}, and received little subsequent attention. It adopts the unique layered crystal structure shown in Fig. \ref{fig:crystal}a that contains tetramers of Cr atoms arranged in a rhombus or lozenge shape. This structure is described in detail in the literature \cite{Klepp-1982, Jobic-1992, Canadell-1992} and in the discussion below. Canadell et al. \cite{Canadell-1992} identified \crte\ as a Mott-Hubbard insulator based on electronic structure calculations and the observation of semiconducting behavior in electrical resistance measurements, but were unable to characterize the magnetic properties due to the presence of strongly magnetic impurities in powder samples.

We have grown single crystals of \crte\ and have studied the structural, thermal, magnetic, and transport properties of the material using x-ray and neutron diffraction, physical property measurements, inelastic neutron scattering, and first principles calculations. We find very anisotropic and unusual thermal expansion below room temperature, with evidence of strong magneto-elastic coupling as in-plane magnetic correlations develop. The compound is easily cleavable, which is clear from handling the crystals and from density functional theory calculations incorporating magnetism and the van der Waals interactions. These calculations predict a band gap of 0.26\,eV for a magnetically ordered state, close to the values determined from electrical resistivity measurements near room temperature. The magnetic susceptibility follows qualitatively the behavior expected for a two-dimensional antiferromagnet. Alternatively, the data can be described by adapting a spin-lozenge model previously developed for other materials containing spin-3/2 tetramers like those found in \crte. Low dimensional magnetic correlations develop above room temperature, and magnetic order is observed in neutron diffraction measurements below T$_N$\,=\,55\,K. Moments are arranged in an up-down-up-down pattern around the perimeters of each lozenge-shaped tetramers, and the long-range magnetic order consists of stripe-like antiferromagnetic arrangement of the tetramers within each layer, with layers stacked antiferromagnetically. Weak anomalies are observed in the heat capacity and magnetic susceptibility at T$_N$. Inelastic neutron scattering reveals dispersive magnetic excitations at low temperature, supporting the presence of non-negligible inter-tetramer interactions. Together the experimental and theoretical results presented here suggest \crte\ to be (1) an interesting material in which to study low dimensional magnetism in bulk crystals, (2) an easily cleavable material for investigating magnetic monolayers or few layer systems, and (3) a candidate semiconducting material for incorporating magnetism into van der Waals heterostructures.

\section{Procedures}

CrTe$_3$ powder was produced by reacting Cr powder and Te shot in evacuated and sealed silica ampoules. The elements were mixed in a 1:3 molar ratio and heated to 700$^{\circ}$C and held for several hours, then cooled quickly to 425$^{\circ}$C and annealed for about 24 hours. The resulting sample was single phase CrTe$_3$ by powder x-ray diffraction, but some ferromagnetic Cr$_5$Te$_8$ was detected by magnetization measurements. This impurity could be reduced in concentration by annealing the powder at similar temperatures in the presence of Te vapor produced by excess Te shot in the same ampoule but physically separated from the sample, or by starting with a Te-rich composition like CrTe$_{3.1}$.

Single crystals were grown by two methods. The first employed a halide flux consisting of a mixture of KCl and AlCl$_3$ in a 45:55 molar ratio (m.p. 220$^{\circ}$C). Working in a helium filled glovebox, about 100 mg of CrTe$_3$ powder produced as described above was added to the mixture of dry salts in a silica tube (9\,mm I.D., 1.5\,mm wall thickness) that was sealed under vacuum without exposure to air. The tube was heated in a horizontal furnace so that one end was held at 450$^{\circ}$C and the other at 425$^{\circ}$C. After several days to one week the furnace was turned off and the tube cooled to room temperature. Platelike crystals were recovered by washing the flux away with water. This method produced mm-sized thin crystals like those show in Fig. \ref{fig:crystal}b. The second crystal growth method used excess Te to grow CrTe$_3$ directly from the melt. Mixtures of Cr and Te containing 1-3 atomic \% Cr were placed in an alumina crucible covered with an alumina frit \cite{Canfield-2016}, and heated to 1050$^{\circ}$C and held for 16 h, cooled to 600$^{\circ}$C and held for 24\,h, then cooled at a rate of 1.5$^{\circ}$C per hour to 455$^{\circ}$C. At this temperature the excess Te was removed by centrifugation. Crystal produced by this method are shown in Fig. \ref{fig:crystal}c and were typically several mm on a side and often grew in thick, blocklike forms with thicknesses up to several mm, but are easily delaminated into thin sheets. Increasing the Cr concentration in the melt to 4\% produced single crystals of Cr$_5$Te$_8$ instead of \crte, consistent with the published phase diagram \cite{PD}.

Powder and single crystal x-ray diffraction were performed using a PANalytical X'Pert Pro MPD powder diffractometer (Cu-K$_{\alpha1}$ radiation) with an Oxford PheniX cryostat and a Bruker SMART APEX CCD single crystal diffractometer (Mo K$_{\alpha}$ radiation). Powder data was analyzed using FullProf \cite{Fullprof}. Single crystal data were analyzed using SADABS for absorption corrections, XPREP for symmetry analysis, and the ShelX suite for structural solution and refinement. Magnetization measurements were conducted using a Magnetic Property Measurement System (Quantum Design). Heat capacity and electrical resistivity measurements were performed with a Physical Property Measurement System (Quantum Design). Electrical contacts were made using silver paste and platinum wires. Heat capacity was measured on single crystal samples, with measurements down to 380\,mK carried out using the Helium-3 Option.

Neutron powder diffraction measurements were performed using the POWGEN instrument at the Spallation Neutron Source (SNS) at Oak Ridge National Laboratory. For these measurements, the powder sample was loaded into 8\,mm diameter vanadium can, and cooled down to 10\,K using a closed-cycle-refrigerator. Two frames with center neutron-wavelengths of 1.333\,{\AA} and 2.665\,{\AA} were used for collecting data over a sufficiently large d-spacing range (0.5\,{\AA} to 10\,{\AA}) at two different temperatures: 10\,K and 300\,K. A temperature dependence study was carried out by collecting data continuously using the 2.665\,{\AA} frame while cooling the sample between 300\,K and 10\,K at a nominal rate of 0.5\,K/min. The data were rebinned into datasets every 10\,K with a nominal counting time of 20 minutes. The structural data was refined using FullProf \cite{Fullprof}. Neutron diffraction study of a single crystal specimen (identical to that used for the magnetization study) was performed using the HB3A 4-circle diffractometer at the High Flux Isotope Reactor. Data were collected using the neutron wavelength of 1.546 {\AA}. Inelastic neutron scattering (INS) measurements were performed using the hybrid spectrometer HYSPEC at the SNS.  HYSPEC is a highly versatile direct geometry spectrometer that combines time-of-flight spectroscopy with the focusing Bragg optics \cite{Hyspec}. The incident neutron beam is monochromated using a Fermi chopper and is then vertically focused by Bragg scattering onto the sample position by highly oriented pyrolytic graphite. For the INS measurements, approximately 5\,g of \crte\ powder was loaded in an aluminum can and placed in an Orange cryostat capable of reaching 1.5\,K. Data was collected with incident energies $E_i$\,=\,50\,meV and 35\,meV, and Fermi chopper frequency of 300\,Hz.

Density functional theory calculations were performed in order to understand the atomic, magnetic, and electronic properties of \crte. The cleavage energy is calculated using the spin polarized vdW-DF-C09 functional \cite{Dion-2004, Thonhauser-2007, Cooper-2010, Thonhauser-2015} with norm conserving psedopotential using the Quantum Espresso package v5.4.0 \cite{Quantum-Espresso}, in which the spin is included in the non-local part of the exchange-correlation functional. For these calculations a model containing six layers per unit cell was constructed that included 192 atoms. The plane wave cut off was set to 80\,Ry and the Brillouin zone sampling used for the supercell was $1\times3\times3$. This approach does not, however, allow automated relaxation and energy minimization, so for the structural optimizations the Vienna Ab-initio Simulation Package (VASP version 5.3.5) \cite{VASP} with the projector-augmented wave (PAW) \cite{PAW} potentials was used. In both codes, 3p$^6$ 4s$^1$ 3d$^5$ for Cr and 5s$^2$ 5p$^4$ for Te were explicitly included as valence electrons in the pseudopotentials.  Bulk \crte\ with  various magnetic configurations were relaxed to get the optimized geometry. The exchange-correlation was approximated with the generalized gradient approximation (GGA) of Perdew-Burke-Ernzerhof (PBE) functionals \cite{PBE} as well as the vdW-DF-optB86b functional \cite{Dion-2004, Thonhauser-2007, Cooper-2010, Klimes-2011} in order to incorporate van der Waals interaction during structural optimization. The Monkhorst-Pack \cite{Monkhorst-Pack} k-point sampling method in the Brillouin zone with a 4x4x4 and a 8x8x8 meshes for ionic and electronic optimization, respectively. The energy cutoff was 500\,eV and the criteria for energy and force convergence are set to be $1\times10^{-4}$\,eV and 0.01\,eV/{\AA}, respectively.

\section{Results and Discussion}

\subsection{Crystal structure and thermal expansion}
\begin{figure}
\begin{center}
\includegraphics[width=3.25in]{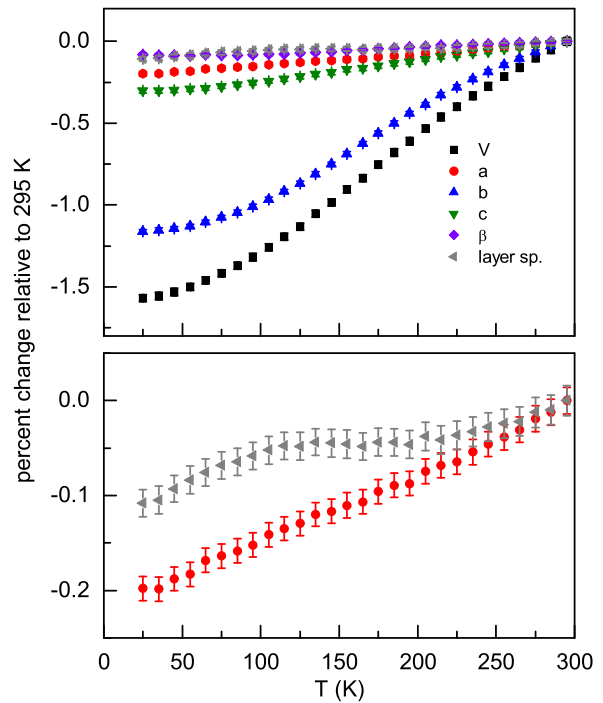}
\caption{\label{fig:lattice}
Percent change in lattice parameters of CrTe$_3$ upon cooling from 295 K determined from powder neutron diffraction data.
}
\end{center}
\end{figure}

Diffraction data collected in this study are consistent with the published crystal structure of \crte\ \cite{Klepp-1982}, and crystallographic parameters determined at several temperatures can be found in the Supplemental Material \cite{SM}. The layered nature of the compound is apparent in Fig. \ref{fig:crystal}a, where the van der Waals gap separating layers of composition \crte\ is visible. A plan view of a layer is also shown. The layers are made up of Cr$_4$Te$_{16}$ units (Fig. \ref{fig:crystal}a) comprising four edge-sharing CrTe$_8$ octahedra. The octahedra share corners with those in neighboring Cr$_4$Te$_{16}$ units to form the two dimensional layers. The four Cr atoms in each unit are arranged in a rhombus or lozenge shape. The intra-lozenge Cr$-$Cr distances labeled by J$_1$ and J$_2$ in Fig. \ref{fig:crystal}a are 3.62$-$3.64 {\AA} and 3.89 {\AA}, respectively, so no Cr-Cr chemical bonding is expected (the Cr$-$Cr distance in Cr metal is 2.5\,{\AA}). There are, however, short distances between Te atoms on neighboring Cr$_4$Te$_{16}$ units, shown by the dark lines in the plan view in Fig. \ref{fig:crystal}a, forming Te$_2$ dimers and Te$_3$ trimers \cite{Klepp-1982}. Such bonding between anions is relatively common in telluride compounds \cite{Jobic-1992}. Simple electron counting gives oxidation states of zero for the Te atoms with two bonds (centers of the trimers), 1- for the Te atoms with one bond (dimers and ends of trimers), and 2- for the Te atoms bonded only to Cr. This results in the expected oxidation state of 3+ for Cr. First principles calculations indicate strong covalency in this material, with nearly equal contribution from Cr and Te in states near the Fermi level, as shown in the Supplemental Material \cite{SM}.

As noted above, crystals of \crte\ are easily separated into thin sheets due to the van der Waals bonding between the layers, suggesting these materials could be of some interest in forming van der Waals heterostructures. To further support and quantify this observation, first principles calculations were performed to estimate the cleavage energy, defined as the energy per unit area required to separate a crystal at the van der Waals gap into two. For the purpose of these calculations, a structural model was built that contained six \crte\ layers per unit cell and the distance ($d$) between two of the layers was gradually increased from the equilibrium value ($d_0$ = 3.5 {\AA}), as shown in Fig. \ref{fig:crystal}d. The figure shows the calculated energy relative to the equilibrium energy as a function of $(d - d_0)/d_0$. The curve approaches a saturation value, the cleavage energy, of about 0.5\,J/m$^2$. This is comparable to well known cleavable materials used for studies of monolayer materials and construction of van der Waals heterostructures, including graphite 0.43\,J/m$^2$ \cite{Bjorkman-2012}, MoS$_2$ 0.27\,J/m$^2$ \cite{Bjorkman-2012}, CrSiTe$_3$ and CrGeTe$_3$ 0.35$-$0.38\,J/m$^2$ \cite{Li-2014}, transition metal thio- and seleno-phosphates (MPS$_3$, MPSe$_3$) 0.35$-$0.55\,J/m$^2$ \cite{ZhangMPS3-2015}, and chromium trihalides 0.3\,J/m$^2$ \cite{McGuire-2015, ZhangCrX3-2015}.

The temperature dependence of the monoclinic lattice parameters, unit cell volume, and layer spacing determined from neutron powder diffraction measurements are shown in Fig. \ref{fig:lattice} with more details in the Supplemental Material \cite{SM}. The layer spacing is defined as the distance between neighboring layers measured perpendicular to the planes of the layers (along the \textit{a}* reciprocal lattice direction) and is equal to a$\cdot$sin($\beta$). Very anisotropic thermal expansion is observed, as is often the case for layered, van der Waals bonded materials, in which the bonding between the layers is weak compared to the intralayer bonding. However, in CrTe$_3$, the thermal expansion is largest along the \textit{b}-axis, which lies within the layers, and the \textit{a}-axis and layer separation have only a weak temperature dependence. This is seen more clearly in the lower panel of Fig. \ref{fig:lattice}.

To investigate the origin of this unusual thermal expansion behavior, full single crystal diffraction datasets were collected at 250, 173, and 100\,K, and refined to determine accurately the atomic positions and interatomic distances. Results of the refinements are included in the Supplemental Material \cite{SM}. Knowledge of the coordinates of each Te atom allows measurement of the van der Waals gap, defined as the perpendicular distance between Te planes on each side of the gap, and the layer thickness, defined as the perpendicular distance between Te planes on each side of a Cr plane (Fig. \ref{fig:crystal}a). The layer spacing as defined above is the sum of these two distances. This analysis shows that the van der Waals gap distance does indeed decrease upon cooling, by 0.2\% between 250 and 100 K, but this is compensated by an increase of nearly the same amount in the layer thickness, resulting in little change in the layer spacing. It is expected that the increase in thickness of the layers upon cooling is driven by the very strong contraction of the in-plane b lattice parameter seen in Fig. \ref{fig:lattice}a. Within the Cr sublattice, the in-plane contraction arises mostly from the contraction of the tetramers themselves, which actually become further separated from neighboring tetramers upon cooling. The Te-Te dimers and Te-Te-Te trimers are quite rigid, with intra-dimer and intra-trimer distances changing by about 0.1\% between 250 and 100 K, while distances between Te atoms not bound into these polymeric units decrease by about 1\%.

In addition to the unusual relative thermal expansion among the different crystallographic directions, anomalous behavior in the temperature dependence of individual lattice parameters is also seen (Fig. \ref{fig:lattice}). This is most apparent in the layer spacing, but the in-plane lattice parameter \textit{b} and unit cell volume \textit{V} also behave in an unusual way. In particular, negative curvature is observed in these data from room temperature down to about 150\,K. These behaviors are expected to be related to the onset of magnetic correlations, since first principles calculations discussed below show notable variation in the optimized lattice parameters when different magnetic configurations are used. Thus the present results indicate strong magneto-elastic coupling is present in \crte.

%---------------------------------LATTICE DISCUSSION END

\subsection{Electrical resistivity}
\begin{figure}
\begin{center}
\includegraphics[width=3.25in]{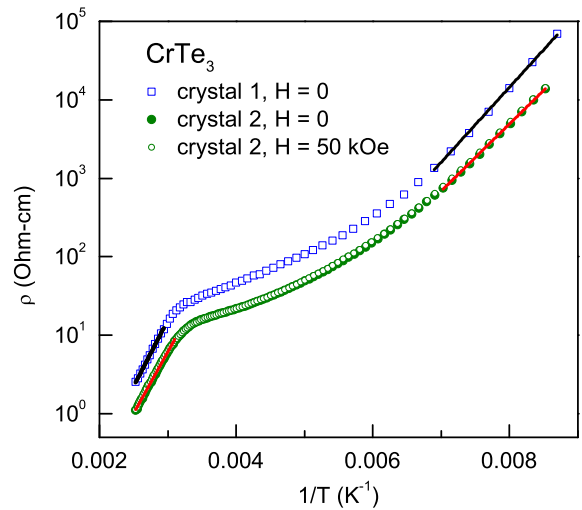}
\caption{\label{fig:res}
Electrical resistivity of a CrTe$_3$ crystal measured in the bc-plane in zero magnetic field and in a magnetic field of 50 kOe directed out of the plane.
}
\end{center}
\end{figure}

The semiconducting nature of CrTe$_3$ is illustrated by the electrical resistivity ($\rho$), which was measured with the current in the bc-plane of two crystals grown using a tellurium flux (Fig. \ref{fig:res}). The same qualitative behavior is seen in both samples, but the absolute values differ by about a factor of 2 over the entire temperature range. This is likely attributable to uncertainty in the actual thickness through which the current flows in this easily delaminated material. Comparison of data collected in zero field and in a 50\,kOe applied magnetic field indicates negligible magnetoresistance. The transport in CrTe$_3$ is clearly activated, with a change in activation energy occurring near room temperature. Linear fits to log($\rho$) vs 1/T at high and low temperatures are shown for both samples in Fig. \ref{fig:res}. From these fits, activation energies of 0.07$-$0.08\,eV and 0.14$-$0.15\,eV are determined at low and high temperatures, respectively. The low temperature behavior likely arises from activation of defects or impurities acting as donors or acceptors. Intrinsic semiconducting behavior may be responsible for the behavior above room temperature. If that is the case, a band gap of 0.28$-$0.30 eV can be estimated. An activation energy of 0.35\,eV has been reported based on resistivity measurements on a cold-pressed polycrystalline sample of CrTe$_3$, but that determination was made using data only below 300\,K \cite{Canadell-1992}. Interestingly, first principles calculations (see \cite{SM}) give a similar gap of 0.26\,eV in the magnetically ordered state.

%---------------------------------RESISTIVITY DISCUSSION END

\subsection{Heat capacity}
\begin{figure}
\begin{center}
\includegraphics[width=3.25in]{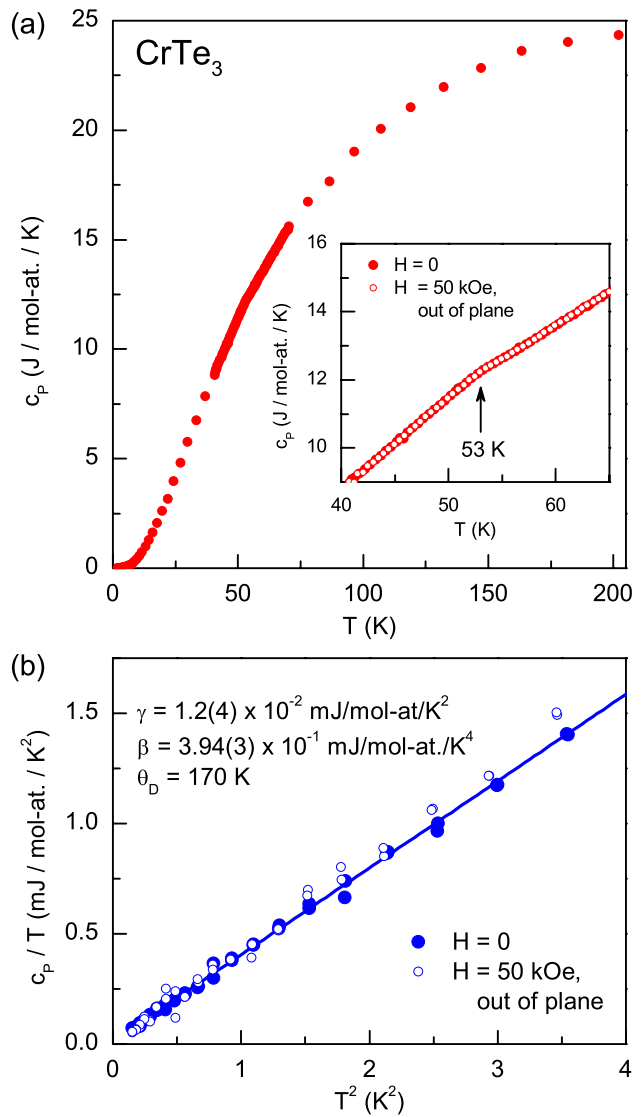}
\caption{\label{fig:hc}
(a) Heat capacity of a \crte\ single crystal. The weak anomaly observed near 53\,K is shown in the inset. (b) Heat capacity between 0.38 and 2\,K collected in zero and 5\,T magnetic fields showing c$_P$ $\propto$ T$^3$ behavior. A linear fit to the zero field data is shown and the resulting parameters are listed on the plot.
}
\end{center}
\end{figure}
The specific heat capacity (c$_P$) of a CrTe$_3$ single crystal is shown in Fig. \ref{fig:hc}. The data reach 24.4\,J/mol-at./K at 200 K (Fig. \ref{fig:hc}a), close to the Dulong-Petit limit of 24.9\,J/mol-at./K. No large anomalies typically associated with long range magnetic order are seen; however, close inspection reveals a weak anomaly that likely indicates a broad lambda-like peak centered at about 53\,K (Fig. \ref{fig:hc}a, inset). Data were collected through this transition in zero field and in an applied field of 50 kOe, and no field effect is observed.

Lower temperature data on a single crystal was also collected to look for additional anomalies. The results are shown in Fig. \ref{fig:hc}b. Debye-like behavior is seen to persist down to 380\,mK. A linear fit to c$_P$/T vs T$^2$ for the data collected in zero field is shown. The \textit{y}-axis intercept of the fit is nearly zero, as expected for a semiconductor with no T-linear electronic contribution to the heat capacity. Assuming the heat capacity in this temperature range is entirely due to phonons, a Debye temperature of $\theta_D$\,=\,170\,K can be determined (although antiferromagnetic magnons would also contribute to the T$^3$ term). This is consistent with the high temperature behavior of c$_P$, which reaches 95\% of the 3R at this temperature. Data collected with a 50\,kOe magnetic field applied perpendicular to the bc-plane is shown in Fig. \ref{fig:hc} as well. The magnetic field is seen to have a negligible effect.

%---------------------------------HEAT CAPACITY DISCUSSION END

\subsection{Magnetic Properties}

\begin{figure}
\begin{center}
\includegraphics[width=3.25in]{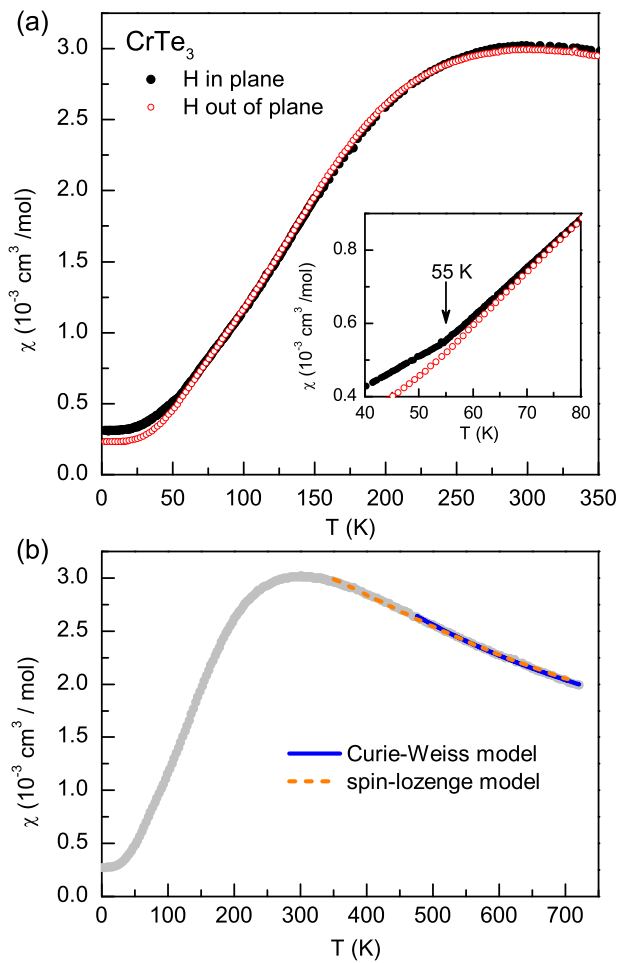}
\caption{\label{fig:mag}
(a) Magnetic susceptibility of CrTe$_3$ measured on a single crystal with a field of 5\,T applied in plane (in the bc plane in which the layers lie) and out of plane (along a*, the stacking direction). The inset highlights the anisotropy developing below about 55\,K. (b) Results of fitting the high temperature magnetic susceptibility data using models described in the text.
}
\end{center}
\end{figure}

Figure \ref{fig:mag}a shows the magnetic susceptibility ($\chi$) measured on a single crystal between 2 and 380\,K with the field along the layer stacking direction (out of plane) and with the field in the plane of the \crte\ layers (in plane). Higher temperature measurements were performed on a sample comprising four single crystals. Figure \ref{fig:mag}b is a composite of that high temperature data (300$-$720\,K) and the powder average from the single crystal measurements shown in Fig. \ref{fig:mag}a. (2$-$380\,K). At higher temperature $\chi$ shows an unusual temperature dependence, with a broad hump at room temperature. A Curie-Weiss model can only describe the data at the highest temperatures (Fig. \ref{fig:mag}b). Fitting of data above 475\,K gives an effective moment of 4.0\,$\mu_B$ per Cr (expect 3.87\,$\mu_B$ for Cr$^{3+}$) and a Weiss temperature of -285\,K, indicating relatively strong antiferromagnetic interactions are present.

Inspection of $\chi$ near the temperature at which the weak heat capacity anomaly was noted above (see Fig. \ref{fig:hc}a) reveals an anomaly in the magnetic susceptibility as well (Fig. \ref{fig:mag}a). Upon cooling below about 55\,K anisotropy develops in the magnetic susceptibility due to a change in slope of the in-plane data. This is highlighted in the inset of Fig. \ref{fig:mag}a. This behavior is interpreted as the onset of magnetic order, as supported by the neutron diffraction results below. A similarly weak anomaly is observed in the magnetic susceptibility of the quasi-1D magnetic material CrSb$_2$ near its Neel temperature \cite{Hu-2007, Sales-2012}. In analogy with CrSb$_2$, which was shown to have strongly one dimensional magnetic interactions \cite{Stone-2012}, it is expected that magnetic correlations in \crte\ developing well above the long range ordering temperature are responsible for the weakness of the heat capacity and magnetic susceptibility anomalies near 55\,K.

The broad hump in $\chi$ is suggestive of low dimensional magnetic correlations. For example the susceptibility data shown in Fig. \ref{fig:mag}a are reminiscent of the quasi-2D antiferromagnet K$_2$V$_3$O$_8$, which shows a broad maximum near 10\,K and long range magnetic order below 4\,K \cite{Liu-1995, Lumsden-2001}, and layered thiophosphates like MnPS$_3$, which shows a broad maximum near 120\,K and long range order below 78\,K \cite{Clement-1980, Wildes-2006}. The interpretation of the magnetic data as arising from 2D correlations is consistent with the layered nature of the crystal structure of \crte; however, since the \crte\ layers themselves are made up of Cr$_4$Te$_{16}$ units containing tetramers of Cr, a zero-dimensional or molecular magnetism viewpoint can also be considered. Magnetism associated with lozenge shaped tetramers formed by spin-$\frac{3}{2}$ magnetic ions like those found in CrTe$_3$ has been studied in several systems \cite{Flood-1969, Kakos-1969,Drillon-1977, Prasad-2016}, by assuming magnetically isolated tetramers with intra-tetramer interactions described by J$_1$ and J$_2$ defined in Fig. \ref{fig:crystal}a. Results of fitting the the magnetic susceptibility of \crte\ using the model of Drillon et al. \cite{Drillon-1977}, assuming g\,=\,2 and a spin of 3/2 per Cr, are shown in Fig. \ref{fig:mag}b. The model works reasonably well down to 350\,K, and gives J$_1$/k$_B$\,=\,-39.9(1)\,K and J$_2$/k$_B$\,=\,35(2)\,K. The model does not work well at lower temperatures, likely due to the other magnetic interactions that are not taken into account. Since long range magnetic order is observed, it is assumed that non-negligible inter-tetramer interactions are present in \crte. The same was noted for Na$_3$RuO$_4$ by the authors of Ref. \citenum{Haraldsen-2009}, and it was concluded that the noninteracting dimer or tetramer approximation is expected to be violated as a result. However, it is interesting to note that the negative J$_1$ and positive J$_2$ determined by the fit would favor an ``up-down-up-down'' spin configuration around the perimeter of the lozenges.

\begin{figure}
\begin{center}
\includegraphics[width=3.25in]{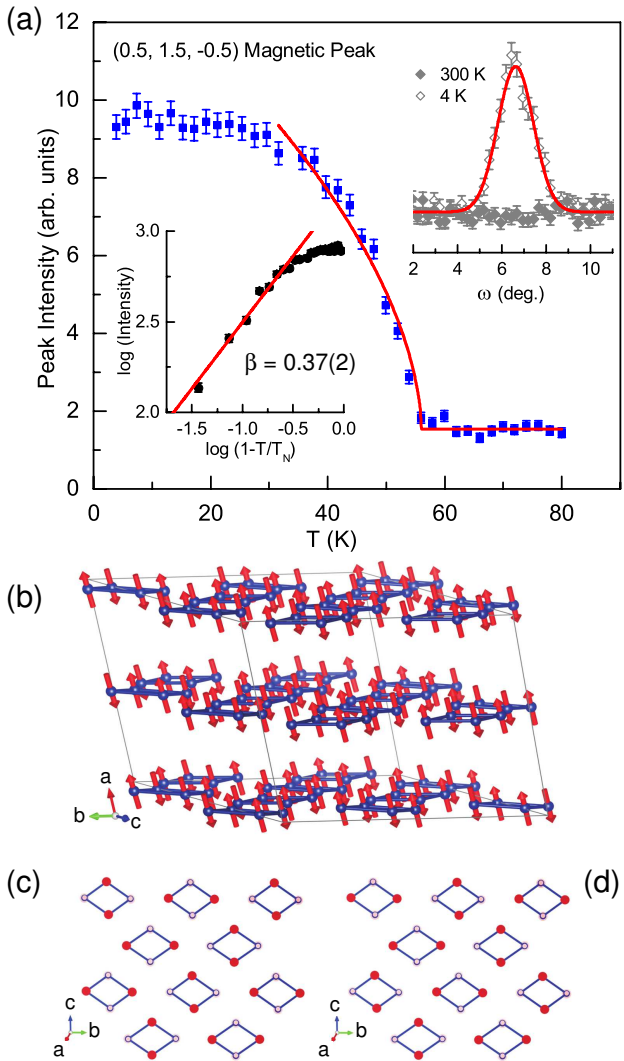}
\caption{\label{fig:neutron}
(a) The order parameter (magnetic Bragg peak intensity) measured using the (0.5, 1.5, -0.5) reflection. The insets show an $\omega$-scan for this reflection measured at 4 and 300\,K, and the fit used to determine the critical exponent $\beta$ using the order parameter data down to 40\,K. (b) The magnetic structure of \crte\ determined from single crystal neutron diffraction at 4\,K. Panels (c) and (d) show two domain configurations within the layers; solid and open circles represent moments directed out of and into the page, respectively. The layers stack antiferromagnetically.
}
\end{center}
\end{figure}

Neutron powder diffraction was used to look for further evidence of magnetic ordering in CrTe$_3$. Comparison of data collected at 300 and 10\,K reveals the presence of additional scattering at 10\,K that is absent at higher temperatures (see Supplemental Material \cite{SM}). No additional scattering was observed in x-ray diffraction between 300 and 20\,K, indicating the feature noted in Fig. \ref{fig:neutron}a is magnetic in nature. The observation of this magnetic Bragg peak, which is centered near Q\,=\,0.945\,{\AA}$^{-1}$ (d\,=\,6.65\,{\AA}), indicates the onset of magnetic order.

Further investigation using a single crystal sample at the HB3A diffractometer allowed identification of the magnetic ordering vector (wavevector) as \textbf{k}=(0.5, 0.5, 0.5) and the ordering temperature $T_N$\,=\,56(1)\,K, which is consistent with magnetization and heat capacity results. Figure \ref{fig:neutron}a shows the order parameter measurement of the (0.5, 1.5, -0.5) magnetic Bragg peak as a function of temperature. The temperature dependence of the peak intensity below $T_N$ was fit to a power law I(T)\,$\propto$\,(1-T/T$_N$)$^{2\beta}$, where $\beta$ is the critical exponent. This is shown on the lower left inset of Fig. \ref{fig:neutron}a. The data obeys this relationship reasonably well down to about 16\,K below T$_N$. A linear fit on the log-log plot in this temperature range gives a value of $\beta$\,=\,0.37(2), which is similar to the theoretical prediction for the 3D Heisenberg model.

To evaluate the possible spin configurations, a magnetic symmetry analysis has been carried out using the tools available at the Bilbao Crystallographic Server \cite{Bilbao}. The best fit of the 43 independent magnetic reflections has been obtained using a magnetic structure model defined by the magnetic space group $P_{S}\overline{1}$ , shown in Fig \ref{fig:neutron}b. In this model, each lozenge-shaped tetramer has an up-down-up-down spin arrangement, with the alternating nonequivalent Cr sites carrying antiparallel spins.  The successive tetramers alternate their spins' directions along all three crystallographic axes, leading to a doubling of the magnetic unit cell with respect to the nuclear cell in all three directions. The nearest neighbor tetramers can be arranged inside the \textit{bc} plane in two equivalent configurations (magnetic domains) that produce stripes aligned  along either the [0,1,1] or [0,-1,1] directions. The two magnetic domains are displayed in Figure \ref{fig:neutron}c and d. Our neutron data indicates an almost equal domain population (0.502(7)\% and 0.498(7)\%). The refined magnetic moment is 2.0(2) $\mu_B$, with the following components along the crystallographic directions: $m_a = 2.1(1)$, $m_b = 0.7(1)$, and $m_c = 1.1(1)\mu_B$. The amplitude of the ordered moment is significantly smaller than the 3\,$\mu_B$ moment expected for trivalent Cr (S\,=\,3/2). Strong covalency between Cr and Te or direct Cr-Cr chemical bonding could reduce the Cr moment; however, electronic structure calculations described below give the full 3.0\,$\mu_B$ moment for Cr. Thus the reduced moment observed by neutron diffraction may be ascribed to incomplete magnetic ordering between the spin-lozenges, with residual disorder most likely along the layer stacking direction. This is supported by the neutron powder diffraction data, in which the magnetic (0.5, 1.5, -0.5) is broader and more asymmetric than the nearby nuclear 100 reflection \cite{SM}. The intra-lozenge order is expected to be complete, as evidenced by the small value of $\chi$ and lack of a Curie tail at low temperatures (Fig. \ref{fig:mag}).

Stability of possible magnetic structures were compared using DFT calculations. Non-magnetic (NM), ferromagnetic (FM), and several antiferromagnetic (AFM) structures were considered. Table \ref{tab:theory} shows the optimized lattice parameters from DFT calculations performed using different magnetic configurations and calculational methods. All calculations used a 2x1x1 supercell model. As expected, neglecting the long range van der Waals interaction (PBE) strongly overestimates the \textit{a}-axis length. Of the magnetic models used, the AFM2 structure is the closest approximation to the experimental structure, which would require a 2x2x2 supercell to represent exactly. In AFM2 (and AFM3) every lozenge within one layer has the same up-down-up-down configuration, while in the experimental structure each layer contains stripes of lozenges with up-down-up-down and down-up-down-up configurations. All of the magnetic structures considered are at least 330\,meV more stable than the NM state. Among the different magnetic configurations, AFM2 and AFM3 are more stable than FM and AFM1 by about 30\,meV. Although this is near the limit of precision for the calculations, it does support the experimentally determined magnetic structure. The calculations give a magnetic moment of 3.0\,$\mu_B$ per Cr. The associated density of states in AFM2 state is included in the Supplemental Material \cite{SM}.

Some variation in the relaxed lattice constants is observed when different magnetic configuration are used in the calculations (Table \ref{tab:theory}). In particular, the in-plane lattice parameters (b and c) are are significantly affected by the inclusion of magnetism. This is identified as an indication of magneto-elastic coupling in \crte, which was also noted above in the temperature dependence of the lattice parameters (Fig. \ref{fig:lattice}).

\begin{table*}
\begin{center}
\caption{\label{tab:theory} Lattice parameters and energies (relative to NM, the non-magnetic state) from DFT calculations using the magnetic configurations and methods listed. For the vdW-DF calculations the OptB86b functional was used.
}
\setlength{\tabcolsep}{2mm}
\begin{tabular}{cccccccc}															
\hline															
configurations	&	description	&	method	&	\textit{a} ({\AA})	&	\textit{b} ({\AA})	&	\textit{c} ({\AA})	&	 $\beta$ (deg.)	&	E-E$_{NM}$ (eV/F.U.)	\\
\hline															
NM	&	non-magnetic	&	vdW-DF	&	7.854	&	10.841	&	11.180	&	117.5	&	0.000	\\
FM	&	ferromagnetic	&	vdW-DF	&	7.888	&	11.507	&	11.696	&	118.9	&	-0.330	\\
AFM1	&	FM layers, AFM stacking	&	PBE	&	8.202	&	11.487	&	11.647	&	117.7	&	--	\\
AFM1	&	FM layers, AFM stacking	&	vdW-DF	&	7.877	&	11.523	&	11.696	&	118.9	&	-0.336	\\
AMF2 	&	AFM layers*, FM stacking	&	vdW-DF	&	7.946	&	11.177	&	11.624	&	118.1	&	-0.363	\\
AFM3	&	AFM layers*, AFM stacking	&	vdW-DF	&	7.935	&	11.178	&	11.623	&	118.1	&	-0.362	\\
\hline
\multicolumn{8}{l}{*layers composed of lozenges with identical up-down-up-down spin configurations, see text.}

\end{tabular}															
\end{center}
\end{table*}

\begin{figure}
\begin{center}
\includegraphics[width=3.25in]{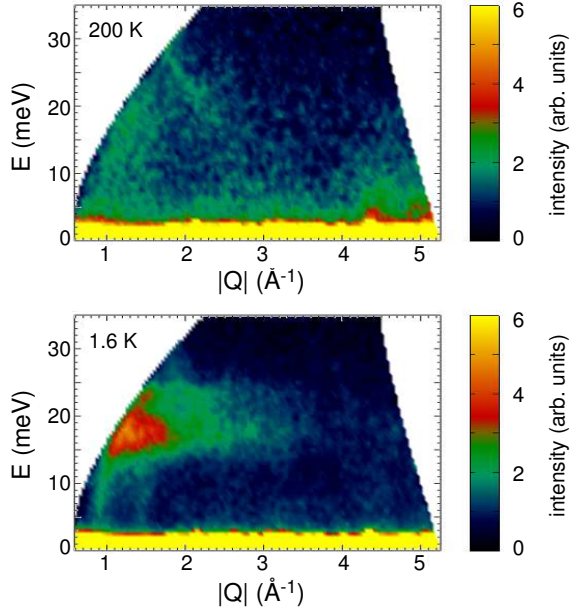}
\caption{\label{fig:INS}
Contour plots of INS intensity from \crte\ powder versus energy and momentum transfer measured at T\,=\,200\,K (upper panel) and 1.6\,K (lower panel). The data were collected using an incident energy of $E_i$\,=\,50\,meV.
}
\end{center}
\end{figure}

Results of inelastic neutron scattering experiments are shown in Fig. \ref{fig:INS}. At 1.6\,K clear evidence of dispersive magnetic excitations are seen that could be associated with the magnetic ordering, which, according to the diffraction data, arises below about 56\,K. The spin-wave-like excitations propagate out of the ordering wave-vector, at Q$\approx$0.945\,{\AA}$^{-1}$ and extend up to approximatively 25\,meV. The inelastic scattering intensity decreases with increasing wave-vector, following the decay of the magnetic form factor, which confirms the magnetic nature of these excitations.  At 200\,K , the intensity of inelastic scattering becomes much weaker and broadens in both energy and momentum space, consistent with an evolution to a paramagnetic state. Phonon scattering becomes more prominent at larger wavevectors above 4\,{\AA}$^{-1}$. For an isolated tetramer scenario, dispersionless excitations would be expected with a Q dependence arising from the dynamic structure factor dependent on intra-tetramer atomic separations. Using the exchange values determined from the magnetic susceptibility, and the calculations for the allowed excitation energies for a spin-3/2 lozenge model given in Eq. 3 of Ref. \citenum{Haraldsen-2009}, one would expect to observe two localized magnetic modes at 3.4\,meV and 6.0\,meV. Instead, dispersive magnetic excitations are observed indicating the presence of significant inter-tetramer couplings and consistent with the long-range ordered ground state. Preliminary analysis indicates that the excitation spectrum can be reproduced using a spin wave model based on three exchange couplings, which includes the two intra-tetramer couplings (J$_1$ and J$_2$) and one antiferromagnetic inter-tetramer interaction (J$_3$). Similar dispersive magnetic excitations have been observed in Na$_3$RuO$_4$ \cite{Haraldsen-2009}, and copper tellurate compounds Cu$_2$Te$_2$O$_5$Cl$_2$ and Cu$_2$Te$_2$O$_5$Br$_2$ \cite{Crowe-2005}, that were previously proposed to consist of coupled spin tetrahedra and exhibiting magnetic long-range order at low temperatures.

%---------------------------------MAGNETIZATION DISCUSSION END

\section{Summary and Conclusions}

\crte\ is an easily cleavable magnetic material with a unique layered crystal structure. The results presented here suggest significant magneto-elastic coupling in \crte\ and reveal low dimensional magnetic correlations extending to above room temperature and long range magnetic order below T$_N$\,=\,55\,K. The structure contains lozenge shaped tetramers of Cr that adopt an up-down-up-down spin arrangement around their perimeters in the magnetically ordered state. There is stripe-like ordering among tetramers within planes and antiferromagnetic stacking between planes. Transport properties show semiconducting behavior consistent with a small band gap well above T$_N$. This may arise from a Mott-Hubbard mechanism since Cr$^{3+}$ has half-filled t$_{2g}$ orbitals, although strong hybridization between Cr and Te is indicated by electronic structure calculations. Based upon the observed electronic and magnetic properties, bulk \crte\ crystals should be of interest for the study of low dimensional magnetism, and the cleavability of the material makes it an exciting candidate for exploring magnetism in monolayer materials and for incorporation into van der Waals heterostructures.

\section*{Acknowledgements}

Research sponsored by the US Department of Energy, Office of Science, Basic Energy Sciences, Materials Sciences and Engineering Division, and Scientific User Facilities Division (neutron scattering). The authors thanks Ben S. Conner and Andrew F. May for helpful discussions through the course of the work, and Radu Custelcean for use of and assistance with the single crystal diffractometer.

%\bibliography{CrTe3-paper}% Produces the bibliography via BibTeX.

%%%%%%%%%%%%%%%%%%%%%%%%%%%%%%%%%%%%%%%%%%%%%%%%%%%%%%%%%%%%%%%%%%%%%%%%%%%%%%%

%%%%%%%%%%%%%%%%%%%%%%%%%%%%%%%%%%%%%%%%%%%%%%%%%%%%%%%%%%%%%%%%%%%%%%%%%%%%%%%


\clearpage

\appendix
\renewcommand\thefigure{S\arabic{figure}}
\renewcommand\thetable{S\arabic{table}}
\setcounter{figure}{0}
\setcounter{table}{0}

\section*{Supplementary Material}

\begin{figure*}
\begin{center}
\includegraphics[width=3.5in]{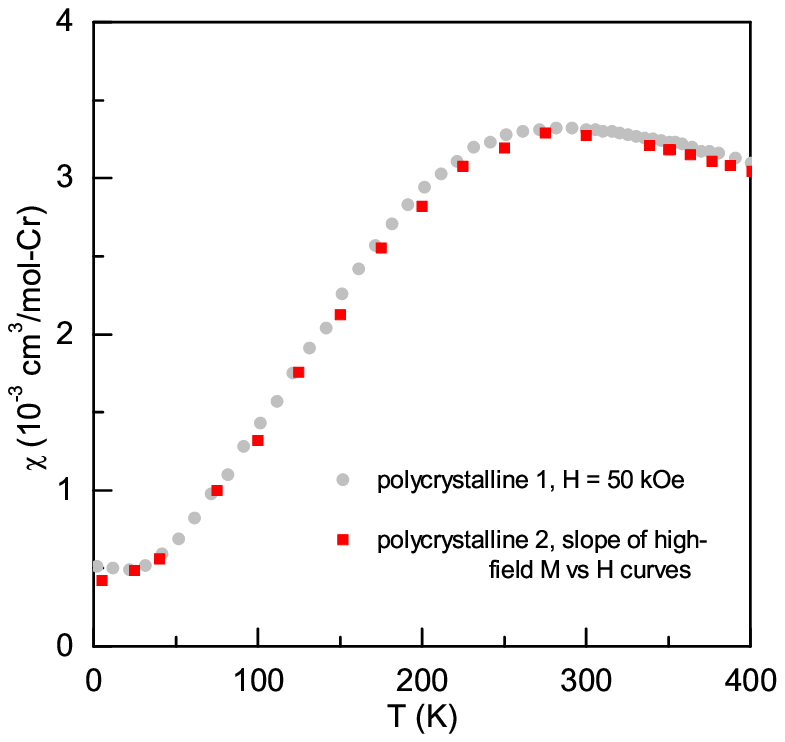}
\caption{\label{fig:poly}
Magnetic susceptibility of two polycrystalline samples of \crte.
}
\end{center}
\end{figure*}

Magnetic susceptibility vs temperature for two polycrystalline samples of \crte\ is shown in Fig. \ref{fig:poly}. Sample 1 is relatively high purity and the data shown are from measurements carried out in a field of 50\,kOe. Sample 2 is less pure, with Cr$_5$Te$_8$ providing a ferromagnetic component below about 280\,K. For this sample the data shown were determined from linear fits to M vs H data between 30 and 50\,kOe.

\begin{figure*}
\begin{center}
\includegraphics[width=5.5in]{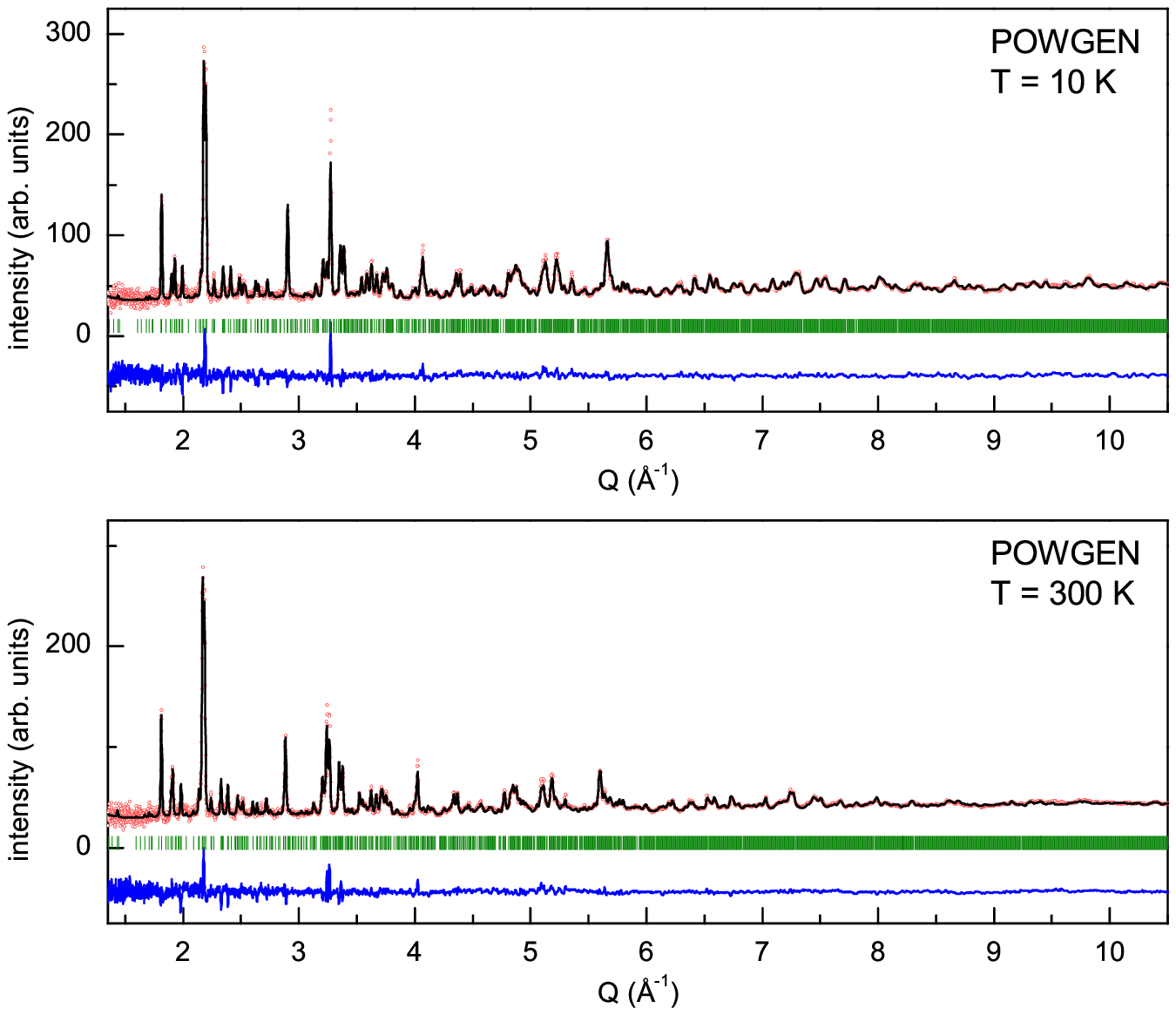}
\caption{\label{fig:fits}
Rietveld fits to neutron powder diffraction data collected at 10 and 300\,K.
}
\end{center}
\end{figure*}

\begin{figure*}
\begin{center}
\includegraphics[width=3.25in]{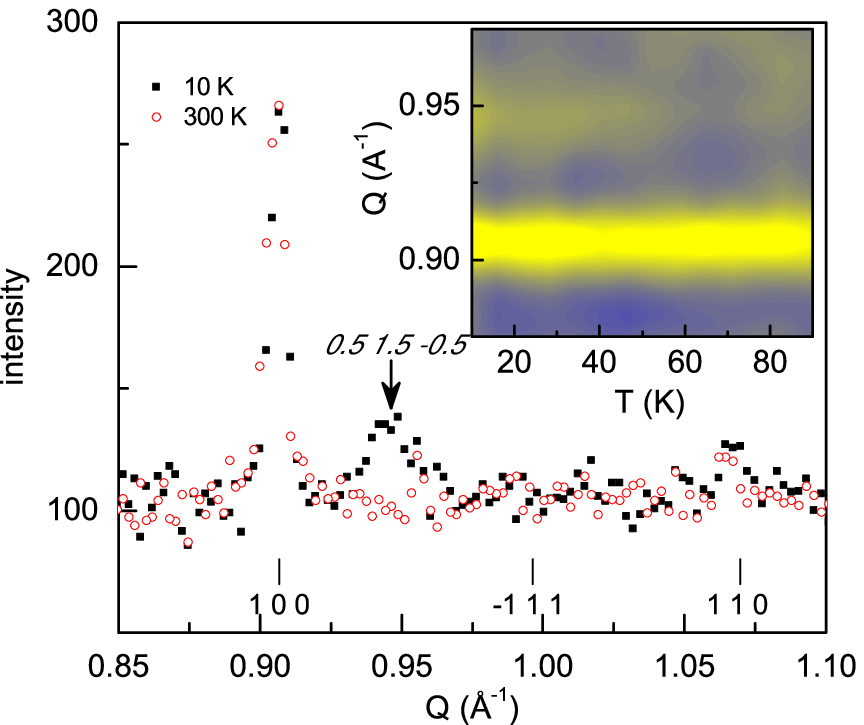}
\caption{\label{fig:powgen}
Powder neutron diffraction data showing a magnetic Bragg peak, marked by the arrow, that appears at low temperature.
}
\end{center}
\end{figure*}

Rietveld fits to neutron powder diffraction data collected at 10 and 300\,K are shown in Figure \ref{fig:fits}. The fitted crystallographic parameters are collected in Table \ref{tab:neut}. Low Q portions of the data are shown in Fig. \ref{fig:powgen}, in which the magnetic Bragg peak (0.5, 1.5, -0.5) can be seen. The limited statistics of this data make determining the ordering temperature difficult, but the contour plot of intensity shown in the inset appears to be consistent with the single crystal data, which showed onset of magnetic scattering near 56\,K. The relative breadth and apparent asymmetry of the magnetic peak suggests that there may be some residual magnetic disorder along the stacking direction at 10\,K.

\begin{figure*}
\begin{center}
\includegraphics[width=5.5in]{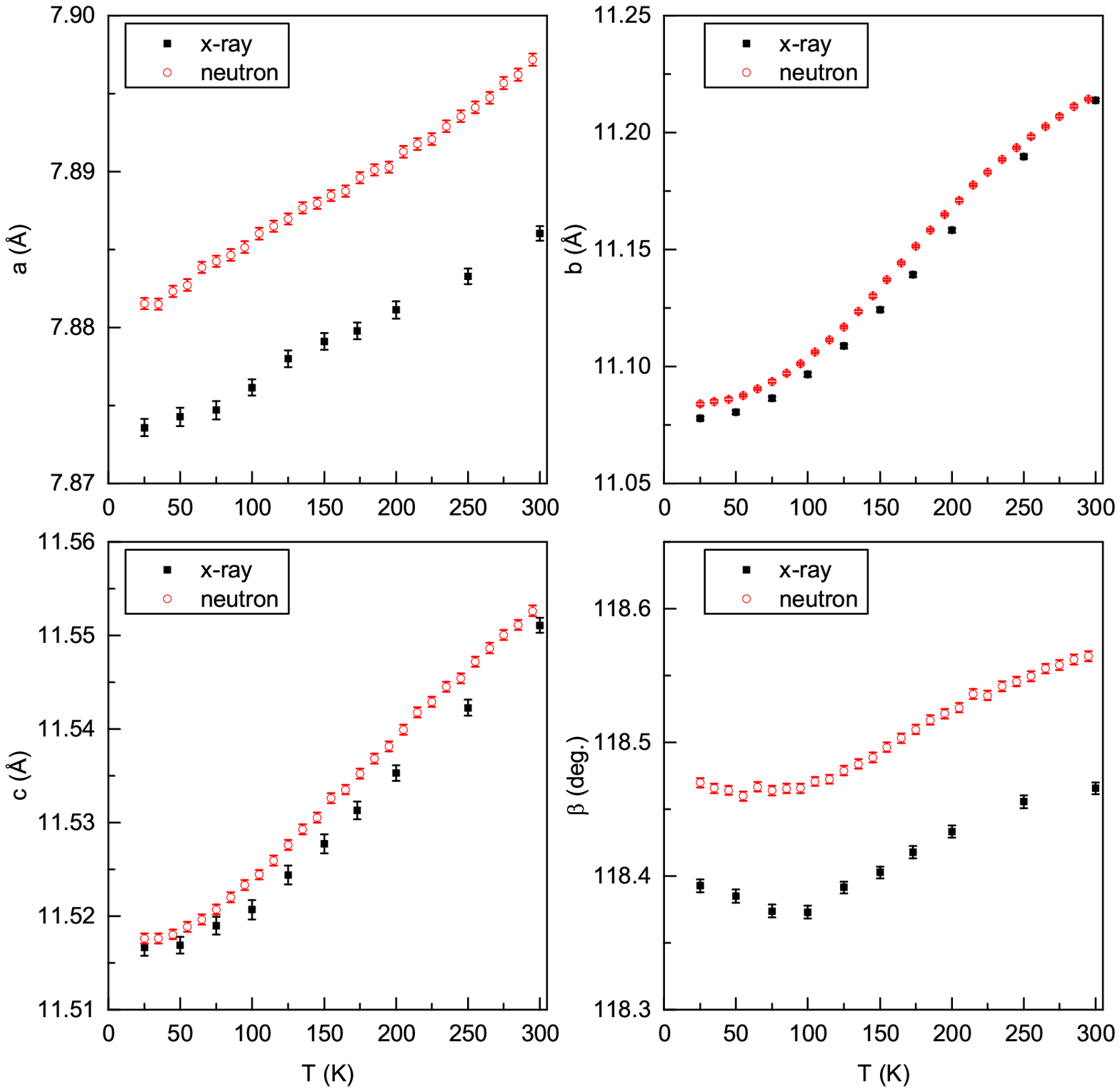}
\caption{\label{fig:compare}
Temperature dependence of the lattice parameters of two CrTe$_3$ samples, one determined by x-ray powder diffraction and the other by neutron powder diffraction.
}
\end{center}
\end{figure*}

The temperature dependence of the lattice parameters determined from neutron and x-ray powder diffraction measurements are shown in Figure \ref{fig:compare}. The largest discrepancy is seen in \textit{a}, which varies by about 0.1\% between the two measurements. This is expected to be due primarily to small sample to sample variations, and it may indicate some degree of stoichiometric variation in the CrTe$_3$ phase. Values for \textit{a} ranging from 7.884 to 7.899\,{\AA} were observed at room temperature based on powder x-ray diffraction measurements.

Results of single crystal x-ray diffraction refinements are collected in Table \ref{tab:x-ray}, which lists lattice parameters and agreement indices, as well as atomic coordinates and equivalent isotropic displacement parameters.

\begin{figure*}
\begin{center}
\includegraphics[width=5in]{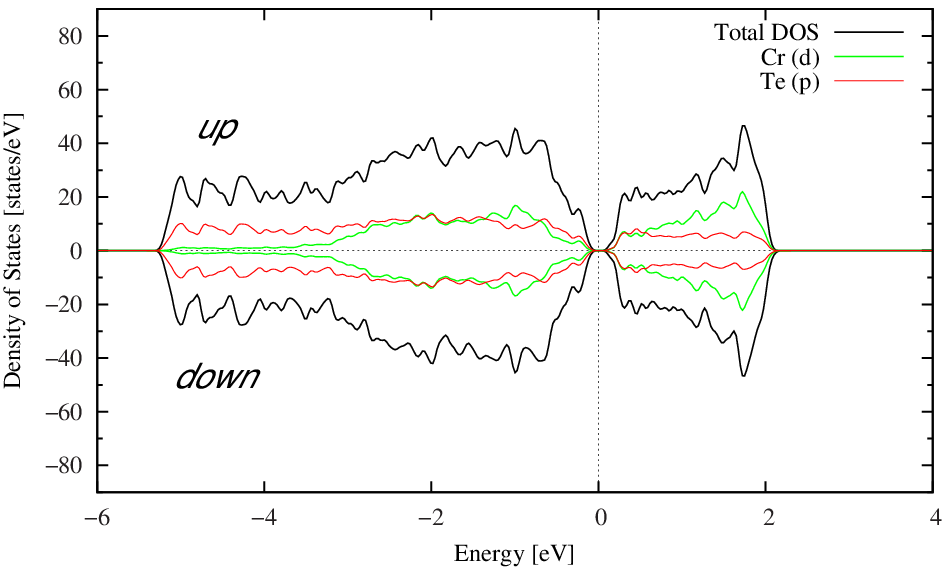}
\caption{\label{fig:DOS}
Calculated total and projected density of states (DOS) for \crte\ in the AFM2 ordered state described in the text. The zero of the energy indicates the Fermi level.
}
\end{center}
\end{figure*}

The calculated total and projected densities of states for \crte\ are shown in Fig. \ref{fig:DOS}. This calculation was performed using the AFM2 configuration.

The positions and moments on the Cr atoms determined by single crystal neutron diffraction refinement are listed in Table \ref{tab:magstr}.

\begin{table*}
\caption{\label{tab:neut} Lattice parameters and agreement indices for CrTe$_3$ using space group $P2_1/c$, and atomic coordinates and isotropic displacement parameters ($B_{iso} = 8\pi^2 U_{iso}$, where $U$ is the mean square atomic displacement) for CrTe$_3$ determined from refinements of powder neutron diffraction data collected at POWGEN.
}
\setlength{\tabcolsep}{8mm}
\begin{tabular}{lll}
\hline					
Temperature	&	300 K	&	10 K	\\
\\					
Unit cell dimensions	&	\textit{a} = 7.896(1) {\AA}	&	\textit{a} = 7.880(1) {\AA}  	\\
	&	\textit{b} = 11.214(1) {\AA}  	&	\textit{b} = 11.084(1) {\AA}   	\\
	&	\textit{c} = 11.552(1) {\AA}	&	\textit{c} = 11.518(1) {\AA} 	\\
	&	 $\beta$ =118.56(1) deg.	&	$\beta$ = 118.46(1) deg.	\\
\\
Volume	&	898.5(2) {\AA}$^3$	&	884.4(1) {\AA}$^3$	\\
\\					
R indices (all data)	&	Rp = 5.25, Rwp = 3.31	&	Rp = 4.89, Rwp = 3.26	\\
\hline					
\end{tabular}

\begin{tabular}{lllll}
\hline									
	&	x	&	y	&	z	&	B$_{iso}$	\\
\hline									
\multicolumn{5}{c}{T = 10 K}									\\
Cr(1)	&	0.008(6)	&	0.503(4)	&	0.170(4)	&	-0.2(3)  {\AA}$^2$	\\
Cr(2)	&	0.022(5)	&	0.233(3)	&	0.008(4)	&	-0.2(3) {\AA}$^2$	\\
Te(1)	&	0.201(5)	&	0.620(2)	&	0.398(3)	&	0.2(2) {\AA}$^2$	\\
Te(2)	&	0.229(4)	&	0.599(2)	&	0.073(3)	&	0.2(2) {\AA}$^2$	\\
Te(3)	&	0.278(3)	&	0.173(3)	&	0.433(2)	&	0.2(2) {\AA}$^2$	\\
Te(4)	&	0.275(3)	&	0.326(3)	&	0.246(2)	&	0.2(2) {\AA}$^2$	\\
Te(5)	&	0.291(4)	&	0.059(2)	&	0.095(3)	&	0.2(2) {\AA}$^2$	\\
Te(6)	&	0.795(5)	&	0.386(2)	&	0.265(3)	&	0.2(2) {\AA}$^2$	\\
\\									
\multicolumn{5}{c}{T = 300 K}									\\
Cr(1)	&	0.011(7)	&	0.503(4)	&	0.175(4)	&	0.3(4) {\AA}$^2$	\\
Cr(2)	&	0.022(5)	&	0.224(3)	&	0.008(4)	&	0.3(4) {\AA}$^2$	\\
Te(1)	&	0.213(6)	&	0.618(2)	&	0.402(4)	&	0.9(3) {\AA}$^2$	\\
Te(2)	&	0.218(5)	&	0.599(3)	&	0.069(4)	&	0.9(3) {\AA}$^2$	\\
Te(3)	&	0.269(4)	&	0.175(3)	&	0.430(3)	&	0.9(3) {\AA}$^2$	\\
Te(4)	&	0.271(4)	&	0.325(3)	&	0.246(3)	&	0.9(3) {\AA}$^2$	\\
Te(5)	&	0.302(4)	&	0.057(3)	&	0.101(4)	&	0.9(3) {\AA}$^2$	\\
Te(6)	&	0.793(5)	&	0.386(3)	&	0.265(3)	&	0.9(3) {\AA}$^2$	\\
\hline									
\end{tabular}

\end{table*}
\begin{table*}
\begin{center}
\caption{\label{tab:x-ray} Lattice parameters and agreement indices, and atomic coordinates and equivalent isotropic displacement parameters from refinements of single crystal x-ray diffraction data from CrTe$_3$ using space group $P2_1/c$ with Mo radiation. The crystal size was approximately 0.070$\times$0.040$\times$0.005 mm$^3$.
}
\setlength{\tabcolsep}{5mm}
\begin{tabular}{llll}
\hline												
Temperature	&	100(5) K	&	173(5) K	&	250(5) K	\\
\\							
Unit cell dimensions	&	\textit{a} = 7.8762(5) {\AA}	&	\textit{a} = 7.8798(5) {\AA}  	&	\textit{a} = 7.8833(5) {\AA}	\\
	&	\textit{b} = 11.0966(8) {\AA}  	&	\textit{b} = 11.1392(8) {\AA}   	&	\textit{b} = 11.1897(7) {\AA}     	\\
	&	\textit{c} = 11.5207(10) {\AA}	&	\textit{c} = 11.5313(10) {\AA} 	&	\textit{c} = 11.5423(9) {\AA}   	\\
	&	 $\beta$ = 118.373(5) deg.	&	$\beta$ = 118.418(4) deg.	&	$\beta$ = 118.456(5) deg.	\\
\\
Volume	&	885.94(12) {\AA}$^3$	&	890.19(12) {\AA}$^3$	&	895.15(11){\AA}$^3$	\\
\\							
R indices (all data)	&	R1 = 0.0336, wR2 = 0.0776	&	R1 = 0.0352, wR2 = 0.0796	&	R1 = 0.0359, wR2 = 0.0781	\\
Largest diff. peak and hole	&	5.780 and -2.434 e/{\AA}$^3$	&	4.592 and -2.894 e/{\AA}$^3$	&	3.510 and -2.749 e/{\AA}$^3$	\\
\hline														
\end{tabular}
\setlength{\tabcolsep}{12.1mm}
\begin{tabular}{lllll}
\hline									
	&	x	&	y	&	z	&	U$_{eq}$	\\
\hline									
\multicolumn{5}{c}{T = 100 K}	\\								
Cr(1)	&	0.0020(1)	&	0.4993(1)	&	0.1680(1)	&	0.007(1) {\AA}$^2$	\\
Cr(2)	&	0.0223(1)	&	0.2297(1)	&	0.0078(1)	&	0.007(1) {\AA}$^2$	\\
Te(1)	&	0.2039(1)	&	0.6206(1)	&	0.3993(1)	&	0.007(1) {\AA}$^2$	\\
Te(2)	&	0.2363(1)	&	0.5918(1)	&	0.0761(1)	&	0.007(1) {\AA}$^2$	\\
Te(3)	&	0.2755(1)	&	0.1683(1)	&	0.4355(1)	&	0.007(1) {\AA}$^2$	\\
Te(4)	&	0.2759(1)	&	0.3273(1)	&	0.2461(1)	&	0.007(1) {\AA}$^2$	\\
Te(5)	&	0.2956(1)	&	0.0561(1)	&	0.0957(1)	&	0.007(1) {\AA}$^2$	\\
Te(6)	&	0.7997(1)	&	0.3799(1)	&	0.2687(1)	&	0.007(1) {\AA}$^2$	\\
\\									
\multicolumn{5}{c}{T = 173 K}		\\								
Cr(1)	&	0.0022(1)	&	0.4991(1)	&	0.1686(1)	&	0.010(1) {\AA}$^2$	\\
Cr(2)	&	0.0221(1)	&	0.2278(1)	&	0.0078(1)	&	0.010(1) {\AA}$^2$	\\
Te(1)	&	0.2049(1)	&	0.6202(1)	&	0.3997(1)	&	0.010(1) {\AA}$^2$	\\
Te(2)	&	0.2342(1)	&	0.5921(1)	&	0.0754(1)	&	0.010(1) {\AA}$^2$	\\
Te(3)	&	0.2739(1)	&	0.1690(1)	&	0.4352(1)	&	0.010(1) {\AA}$^2$	\\
Te(4)	&	0.2743(1)	&	0.3264(1)	&	0.2454(1)	&	0.010(1) {\AA}$^2$	\\
Te(5)	&	0.2970(1)	&	0.0553(1)	&	0.0963(1)	&	0.011(1) {\AA}$^2$	\\
Te(6)	&	0.7985(1)	&	0.3803(1)	&	0.2682(1)	&	0.011(1) {\AA}$^2$	\\
\\									
\multicolumn{5}{c}{T = 250 K}		\\								
Cr(1)	&	0.0022(1)	&	0.4992(1)	&	0.1693(1)	&	0.014(1) {\AA}$^2$	\\
Cr(2)	&	0.0217(1)	&	0.2258(1)	&	0.0076(1)	&	0.013(1) {\AA}$^2$	\\
Te(1)	&	0.2061(1)	&	0.6198(1)	&	0.4002(1)	&	0.014(1) {\AA}$^2$	\\
Te(2)	&	0.2322(1)	&	0.5926(1)	&	0.0748(1)	&	0.013(1) {\AA}$^2$	\\
Te(3)	&	0.2722(1)	&	0.1696(1)	&	0.4350(1)	&	0.014(1) {\AA}$^2$	\\
Te(4)	&	0.2726(1)	&	0.3258(1)	&	0.2446(1)	&	0.014(1) {\AA}$^2$	\\
Te(5)	&	0.2984(1)	&	0.0545(1)	&	0.0969(1)	&	0.014(1) {\AA}$^2$	\\
Te(6)	&	0.7973(1)	&	0.3806(1)	&	0.2676(1)	&	0.014(1) {\AA}$^2$	\\
\hline									
\end{tabular}
\end{center}
\end{table*}

\begin{table*}[]
\centering
\caption{Symmetry operators of Wyckoff positions and the corresponding magnetic moment configuration for the two non-equivalent Cr atoms located inside the crystallographic unit cell. Note the components of the moment along the crystallographic axes are $m_a = 2.1(1)$, $m_b = 0.7(1)$, and $m_c = 1.1(1)\mu_B$. The magnetic structure is defined by the propagation vector \textbf{k}=(0.5,0.5,0.5) and is described by the magnetic space group $P_{S}\overline{1}$.}
\label{tab:magstr}
\begin{ruledtabular}
\begin{tabular}{cc|cc}
%\multicolumn{2}{c}{Cr1~~(0.008,0.503, 0.170)} & \multicolumn{2}{c}{Cr2~~(0.022, 0.233, 0.008)} \\ \hline
Cr1~~(0.008,0.503, 0.170) & & Cr2~~(0.022, 0.233, 0.008) &\\ \hline
$(x, y, z)$ & $(m_a, m_b, m_c)$ & $(x, y, z)$ & $(-m_a, -m_b, -m_c)$  \\
$(-x+1, y-1/2, -z+1/2)$ & $(m_a, m_b, m_c)$ & $(-x+1, y+1/2, -z+1/2)$ & $(m_a, m_b, m_c)$  \\
$(-x+1, -y+1, -z+1)$ & $(m_a, m_b, m_c)$ & $(-x+1, -y+1, -z+1)$ & $(-m_a, -m_b, -m_c)$  \\
$(x, -y+3/2, z+1/2)$ & $(m_a, m_b, m_c)$ & $(x, -y+1/2, z+1/2)$ & $(m_a, m_b, m_c)$   \\
\end{tabular}
\end{ruledtabular}
\end{table*}

\end{document}